\documentclass[twocolumn,prc,showpacs,a4]{revtex4}
\usepackage{psfig,epsfig,graphicx,float,pst-all}
\usepackage{amssymb}
\usepackage{pifont}
\renewcommand{\=}{~=~}
\newcommand{\be}{\begin{equation}}
\newcommand{\ee}{\end{equation}}
\newcommand{\bc}{\begin{center}}
\newcommand{\ec}{\end{center}}

\begin{document}

\title{Solution of the Bohr hamiltonian for soft triaxial nuclei}
\author{L. Fortunato\footnote{ Present address: Dip. di Fisica ``G.Galilei''
, Universit\`a di Padova and INFN, via Marzolo, 8, I-35131, Padova (Italy).\\
 E-mail: {\tt fortunat@pd.infn.it} }, 
S. De Baerdemacker, K. Heyde}
\affiliation{Vakgroep subatomaire en stralingsfysica, Proeftuinstraat 86,
B-9000, Ghent (Belgium)}

\begin{abstract}
The Bohr-Mottelson model is solved for a generic soft triaxial nucleus, 
separating the Bohr hamiltonian exactly and using
a number of different model-potentials: a displaced harmonic oscillator in 
$\gamma$, which is solved with an approximated algebraic technique, and 
Coulomb/Kratzer, harmonic/Davidson and infinite square well potentials in
$\beta$, which are solved exactly. In each case we derive analytic
expressions for the eigenenergies which are then used to calculate 
energy spectra.

Here we study the chain of osmium isotopes and we compare
our results with experimental information and previous calculations.

\end{abstract}
\today
\maketitle

\section{Introduction}
The Bohr-Mottelson collective model of the nucleus has recently attracted 
a significant interest because of the possibility to derive 
many ``new'' solvable 
cases \cite{Iac1,Iac2,Iac3,RoBa,Cap,FV,FV2,Pepe,Bona1,Bona2,BonZ5,Mark2,
Rowe-new,LF,Pietr,CapNew,
RoI,RoTuRe}.
Intense efforts did arise because of the availability of models, based on the 
square well potential, that are related with the issue of critical point 
symmetries at the shape phase transition (E(5), X(5) and Y(5)) \cite{Iac1,
Iac2,Iac3}.
This has given rise from one side to many applications aimed at the survey of 
existing experimental spectroscopic data and at the identification of 
signatures for the new models \cite{Cast,Kru,Capal,Clark,Tonev,Liu,Frank} 
in various mass regions, especially in connection with the effort to 
build new descriptions for transitional nuclei.
On the other side a number of mathematical solutions of the Schr\"odinger 
equation associated with the Bohr hamiltonian with various model-potentials
have been proposed \cite{Cap,FV,FV2,Pepe,Bona1,Bona2,BonZ5,LF,Pietr,CapNew,
RoI,RoTuRe}. 
For some of those potentials, that in general are functions of the quadrupole
deformation variables $\beta$ and $\gamma$, an exact separation of variables
is possible, while in other cases an approximate separation holds. 
Usually the spectrum (and transition rates) is derived analytically and 
compared with available experimental data. We like to mention that
for a number of these new solvable cases extensive comparisons with 
experimental data are still not available. 
 
Recently a solvable model was proposed for the soft triaxial rotor with
a minimum in the potential along the $\gamma$ direction located at $\pi/6$
\cite{LF}.
The aim of the present paper is to extend and complete that solution, 
proposing a solvable model for which the Schr\"odinger equation is separable.
The $\gamma-$angular part is then solved for a harmonic potential centered around any $\gamma_0$ in the 
interval $~]0,\pi/3[~$, while the $\beta$ part is solved exactly for the 
Kratzer-like potential, for the Davidson potential 
and for the infinite square well potential. 

Another reason for the search of analytic solutions for the more general
case with non-irrotational moments of inertia comes also from a recent 
study \cite{Wood} in which the rigid triaxial rotor model is improved 
by relaxing the assumption of irrotational flow moments of inertia. The
fit of the three components of the moment of inertia to experimental data
improves the description of a number of nuclear properties and suggests
that the irrotational assumption is not correct. Our model incorporates
this idea and extends the rigid model to $\gamma$-soft models.

\begin{figure}[!t]
\begin{center}
\epsfig{file=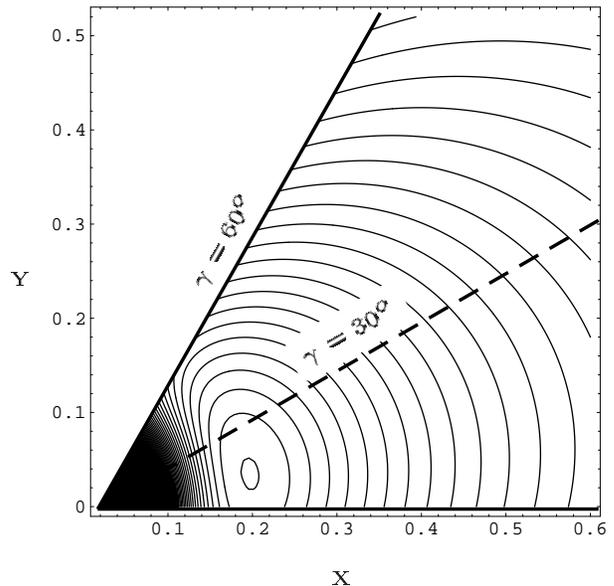,width=8cm,clip=}
\end{center}
\caption{Contour plot of the potential $V(\beta,\gamma)$ with minimum
in $\gamma_0=10^o$ and $\beta_0=0.2$. The situation depicted here 
corresponds to a very soft deformed triaxial rotor, where the $V_1(\beta)$
potential (see Eq. (\ref{pote})) has a Kratzer-like form, while the 
$V_2(\gamma)$ is a displaced harmonic oscillator. 
The coordinates are $X=\beta\cos{\gamma}$ and $Y=\beta\sin{\gamma}$. 
In the point $\beta=0$ the potential goes to $+\infty$ (black area).}
\label{soft10}
\end{figure}

To give an idea of the typical problem that we are considering
we show as an illustration in Fig. (\ref{soft10}) a contour plot of the 
potential surface with a Kratzer-like potential in $\beta$ (with the minimum 
at $\beta_0=0.2$) and a displaced harmonic oscillator potential in $\gamma$ 
(with the minimum at $\gamma_0=10^o$). The solution of the two-dimensional 
Schr\"odinger equation for the Bohr hamiltonian containing this potential
surface is given by a set of eigenvalues and their corresponding eigenstates. 
To each eigenstate (and in particular to the ground state) an average 
quadrupole shape is associated. In the present case (Fig. \ref{soft10}) 
the wave functions are concentrated around the region of the minimum and 
thus we are dealing with a soft triaxial rotor.

The present study may also be considered as an attempt to solve a simple 
model that may be compared with microscopic theories (like HF, HF+BCS or HFB) 
\cite{GG,Hay,Ansari}. 
The potential surfaces used throughout the paper, rather than calculated 
with a variational procedure in a microscopic framework, are 
instead {\it a priori} given in 
analytical form. A number of early microscopic studies, like for example 
Ref. \cite{GG}, concluded that $\gamma-$softness was not only unavoidable 
in triaxial nuclei, but that the potential barrier in the $\gamma$ direction
was so small to cast doubts on the applicability of the rigid rotor model. 
We therefore reanalyze a number of osmium isotopes, which were treated within 
a HFB formalism, and compare our results with the results of Ref. 
\cite{Ansari}. 

In the present paper, we use a potential of the form $V_1(\beta)+V_2(\gamma)/
\beta^2$, which allows for a separation of the Bohr hamiltonian in $\beta$
and $(\gamma,\theta_i)$ part (Section II). The solution of the 
$(\gamma,\theta_i)$ part is amply presented in Section III. 
In Section IV we discuss the quantum numbers associated with the present model
and presents a few comments on quasidynamical symmetry. In Section V we
discuss in detail some examples that illustrate the method of solving Eq. 
(\ref{coupled}).
In Section VI, we solve analytically the $\beta$ part of the problem in the 
Coulomb/Kratzer, harmonic/Davidson and infinite square well cases.
In Sections VII and VIII, we discuss the fitting procedure and its 
accuracy and we apply the method to the chain of osmium isotopes and 
make a comparison with other calculations. Here, we also stress the fact 
that, in doing so, in general we go beyond an irrotational approach
and make use of three different moments of inertia used as parameters.
The need to take into account this possibility has been emphasized before 
by Wood {\it et al.} \cite{Wood}. We also compare with the situation of irrotational
flow in which a value $\gamma_0$ can be deduced and implies that fluctuations 
of the moment of inertia in the $\gamma$ direction are neglected, but
the softness is included. It is interesting to notice that the values of 
$\gamma_0$ so obtained correlate well with results extracted from different 
methods used to analyze data in the Os nuclei (Section VIII).
We analyze these isotopes, particularly because they are thought 
to be situated in the transitional region between $\gamma-$rigid and
$\gamma-$soft shapes. In Section IX, we formulate the main conclusions 
of the present study, while in appendix A, we explicitly present the
matrix elements used in the calculations.

\section{Formulation of the model}
The Schr\"odinger equation for the Bohr hamiltonian reads
\be
\hat H_B \Psi(\beta,\gamma, \theta_\iota)\=
E  \Psi(\beta,\gamma, \theta_\iota)\,,
\ee
where the hamiltonian is given by
\begin{widetext}
\be
\hat H_B\=  -  { \hbar^2\over 2B_m} {1\over \beta^4}{\partial \over 
\partial \beta}\beta^4{\partial \over \partial \beta} -
{\hbar^2\over 2B_m} {1\over \beta^2}{1\over \sin{(3\gamma)}} 
{\partial \over \partial \gamma}
\sin{(3\gamma)}{\partial \over \partial \gamma}+
{\hbar^2\over 8B_m \beta^2}\sum_{i=1,2,3} {{\hat L}_i^2 \over 
\sin^2(\gamma-2\pi\,i/3)} +V(\beta,\gamma)\,.
\ee
\end{widetext}
Here ${\hat L}_i$ are the projections of the angular momentum 
($\hat L$) on the body-fixed axes and $B_m$ is the mass parameter.
An exact separation of the variables $\beta$ and $\gamma$ may be 
achieved when the potential is chosen as 
\be
V(\beta, \gamma)\= V_1(\beta)+{V_2(\gamma)\over \beta^2}\,.
\label{pote}
\ee
The resulting set of differential equations (one containing only the $\beta$
variable and the other containing the $\gamma$ variable and the Euler angles,
$\theta_\iota$ with $\iota=1,2,3$),
after multiplication by $2B_m/\hbar^2$, reads
\be
\Bigl\{-{1\over \beta^4}{\partial \over 
\partial \beta}\beta^4{\partial \over \partial \beta}+u_1(\beta)-
\epsilon+{\omega\over \beta^2}\Bigr\}\xi(\beta)= 0
\label{be}
\ee
$$
\Bigl\{-{1\over \sin{(3\gamma)}} 
{\partial \over \partial \gamma}
\sin{(3\gamma)}{\partial \over \partial \gamma}+\sum_{i=1,2,3} 
{{\hat L}_i^2 \over 4\sin^2(\gamma-2\pi\,i/3)}$$
\be
+u_2(\gamma)-\omega
\Bigr\}\psi(\gamma,\theta_\iota)\=0\,,
\label{g-a}
\ee
where $\omega$ is the separation constant, $\epsilon=(2B_m/\hbar^2)E$
 and $u_t=(2B_m/\hbar^2)V_t$ with 
$t=1,2$. The wavefunctions satisfy $ \Psi(\beta,\gamma, \theta_\iota)\=
\xi(\beta)\psi(\gamma,\theta_\iota)$.
In Ref. \cite{LF} the same derivation has been proposed, but the problem 
was restricted to deriving the solution of the soft triaxial rotor around 
$\gamma\sim\pi/6$. In that case the rotational kinetic term may be
written in a very simple way and the solution of $\gamma-$angular part
may be given in a straightforward way. The main result of Ref. 
\cite{LF} was, (i) to extend the Meyer-ter-Vehn formula 
\cite{MTV} (strictly valid for a rigid rotor at $\gamma=30^o$) proving 
that the corresponding formula for the $\gamma-$soft rotor requires the
addition of a (trivial) harmonic term in the $\gamma$ degree of freedom 
and (ii) to show that the solution of the full problem
does not necessarily imply a trivial extension with some additive term 
to include the $\beta-$vibrations, but rather yields an expression 
for the energy levels in which the quantum numbers are intertwined in a 
more complicated way.

Here we extend the results of Ref. \cite{LF}, addressing the more general 
problem of a soft triaxial rotor (not confined to $\gamma\sim30^o$), 
solving approximatively the $\gamma-$angular equation with a potential
of the form
\be
u_2(\gamma)=C(\gamma-\gamma_0)^2=Cx^2\,,
\label{u2}
\ee
that has a minimum in the interval $0^o<\gamma_0< 60^o$. 
For symmetry reasons we can 
restrict ourselves to the sector $0^o<\gamma_0 \le 30^o$.
The obstacle to the further separation of variables in Eq. (\ref{g-a}) 
is represented by the rotational term (second term), that mixes the 
variable $\gamma$ with the projections of the angular momentum. 
This term can be rewritten as $\sum_i A_i \hat L_i^2$.
We will follow two strategies to deal with the coefficients $A_i$. 
\begin{itemize}
\item{
The first strategy is to approximate these coefficient as follows 
\be
A_i={1\over 4 \sin^2(\gamma_0-2\pi\,i/3)}\,,
\label{a-moment}
\ee
and to replace the corresponding terms in Eq. (\ref{g-a}).
The physical meaning of this approximation is that the 
fluctuations of the moments of inertia are completely neglected, while
the softness is taken into account. Making use of the form (\ref{a-moment}),
we can parametrise the components of the moment of inertia by 
one single parameter.}
\item{
It has been shown in an extension of the rigid rotor model \cite{Wood} 
that the components of the moment of inertia fitted to experimental data 
do not necessarily agree with the irrotational approximation. 
We will therefore include this observation in our model, as 
a second strategy, taking the components of the moment of inertia as
parameters.}
\end{itemize}

In the spirit of finding a simple solution, we 
introduce in Eq. (\ref{g-a}) the further simplification 
$~\sin{3\gamma}\sim  \sin{3\gamma_0}~$, obtaining
\be
\Bigl\{ \overbrace{-{\partial^2 \over \partial x^2}+Cx^2}^{\hat D_2}-\omega
+\sum_{i=1,2,3} A_i{\hat L}_i^2 \Bigr\}\psi(x,\theta)=0\,,
\ee
where the variable $x$, introduced in (\ref{u2}), has been used and  
where the second order differential operator $\hat D_2$ has been defined.
The wavefunction may be written in the unsymmetrized form 
as an expansion in terms of rotational wavefunctions, namely
\be
\psi_{L,M}(x,\theta)\=\sum_{K'} a_{K'}^L(x)
{\cal D}^L_{M,K'}(\theta) \,.
\label{expa}
\ee
where $K'$ is introduced to distinguish the multiple occurrence of 
states with the same $L$. One may notice that the rotational basis has an
SO(2) symmetry reflected in the good quantum number $K'$. 
We now multiply to the left by ${\cal D}^L_{M,K}$ and integrate over the
Euler angles, exploiting the orthonormality property of the Wigner functions
where possible. The result is the set of equations (one for each allowed value
of $K$)
\be
({\hat D_2}-\omega) a^L_K(x) +\sum_{i}A_i\sum_{K'} 
\langle LMK \mid 
\hat L_i^2\mid LMK'\rangle  a^L_{K'}(x)=0\,,
\label{coupled}
\ee
each of which contains matrix elements of the squared components of the 
angular momentum (see appendix). We may rewrite it also as
\begin{widetext}
$$\Bigl({\hat D_2}-\omega+\sum_i A_i\langle LMK \mid 
\hat L_i^2\mid LMK\rangle \Bigr) a^L_K(x)+$$
\be
{(A_1-A_2)\over 4}\langle LMK \mid 
\hat L_+^2\mid LMK-2\rangle a^L_{K-2}(x)+{(A_1-A_2)\over 4}\langle LMK \mid 
\hat L_-^2\mid LMK+2\rangle a^L_{K+2}(x)=0
\label{undici}
\ee 
\end{widetext}
to highlight that only three of the $a$ coefficients appear in every equation.

\section{Algebraic approach}
A convenient way to treat the $\gamma-$angular part is to consider the 
dynamic symmetry associated with the $\hat D_2$ operator, that is essentially 
a harmonic oscillator hamiltonian. The advantage here lies in the fact 
that the spectrum generating algebra allows us to write explicit 
expressions for the $\gamma-$vibrational spectrum.

The infinitesimal generators of the sp(2,$\mathbb{R}$) Lie algebra
may be written as
\be
\hat Z_1=-{1\over k}{\partial^2\over \partial x^2} \qquad \hat Z_2=k x^2 
\qquad \hat Z_3=-i\Bigl({1\over 2} +x{\partial\over \partial x}\Bigr)\,,
\label{op}
\ee
where $k$ is a constant. Similar approaches 
have been used in Ref. \cite{RoBa,FV}. 
The operators above close under commutation, i.e.:
\be
[\hat Z_1,\hat Z_2]=-4i \hat Z_3 \quad
[\hat Z_3,\hat Z_2]=-2i \hat Z_2 \quad
[\hat Z_3,\hat Z_1]=2i \hat Z_1 \,.
\ee
With the linear transformation
\be 
\hat X_1\={1\over 4} \bigl( \hat Z_1-\hat Z_2 \bigr) \qquad ~
\hat X_2\={1\over 2} \hat Z_3 \qquad ~
\hat X_3\={1\over 4} \bigl( \hat Z_1+\hat Z_2 \bigr)\,,
\label{tra1}
\ee
one may recognize the standard commutation relations of the four isomorphic 
Lie algebrae su(1,1)$\sim$so(2,1)$\sim$sp(2,$\mathbb{R}$)$\sim$
sl(2,$\mathbb{R}$) 
\be
[\hat X_1,\hat X_2]=-i \hat X_3 \qquad ~
[\hat X_2,\hat X_3]=i \hat X_1 \qquad ~
[\hat X_3,\hat X_1]=i \hat X_2.
\ee 
It is also very useful to define raising, lowering and weight operators
for this algebra,
\be
\hat X_{\pm} = \hat X_1 \pm i \hat X_2 \qquad
\hat X_0 = \hat X_3\,,
\label{tra2}
\ee
that obey the following commutation relations
\be
[\hat X_+,\hat X_-]=-2\hat X_0 \qquad [\hat X_0,\hat X_\pm]= \pm \hat X_\pm
\,.
\ee
The action of the above operators on orthonormal basis states for the irreps 
of su(1,1) ($\mid n,\lambda \rangle$ with $n=0,1,2,..$) 
is given by the equations:
\begin{eqnarray}
&\hat X_+&\mid n\lambda\rangle =\sqrt{(\lambda+n)(n+1)}\mid n+1,\lambda\rangle
 \\
&\hat X_-&\mid n+1,\lambda\rangle =\sqrt{(\lambda+n)(n+1)}\mid n\lambda\rangle
\\
&\hat X_0& \mid n\lambda\rangle ={1\over 2}(\lambda+2n)\mid n\lambda\rangle \,.
~~~~~~~~~~~~~~~\label{diago}
\end{eqnarray}
The values of $\lambda$ are found by comparing the standard 
eigenvalue 
equation for the Casimir operator of su(1,1) with the eigenvalue equation 
for the same operator, but explicitly realized in the terms of the operators 
defined in (\ref{op}):
\be
\hat \mathbb{C}_2 \mid n\lambda\rangle \= 
\Bigl(\hat X_3^2-\hat X_1^2-\hat X_2^2 \Bigr)\mid n\lambda\rangle\=
{1\over 4}\lambda(\lambda-2)\mid n\lambda\rangle\,.
\ee
The action of the Casimir operator on a given function $\phi(x)$ gives
\be
\hat \mathbb{C}_2 \mid \phi\rangle= 
\Biggl\{ {1\over 8}\Bigl( \hat Z_1\hat Z_2+  
\hat Z_2\hat Z_1\Bigr)-
{\hat Z_3^2 \over 4} \Biggr\}\mid \phi\rangle = 
-{3 \over 16}\mid \phi \rangle\,.
\ee
Therefore we obtain $\lambda={1/2,3/2}$. 
As a consequence of the definitions given in this and in the 
previous section, we notice that  
\be
\hat D_2\=\sqrt{C}(\hat Z_1+\hat Z_2)\=\sqrt{C}4\hat X_0\,,
\ee
where the constant $k$ is defined as $k=\sqrt{C}$. The action of the operator 
$4\hat X_0$ is given, for each allowed value of $\lambda$, by $(1+4n)$
and $(3+4n)$ respectively. Combining the two results we may write 
that the eigenvalue equation for the $\hat D_2$ operator is given by 
\be
\hat D_2 \mid n_\gamma \rangle \= \sqrt{C}(1+2n_\gamma) 
\mid n_\gamma \rangle \,,
\ee
where we dropped the index $\lambda$ to cover the whole spectrum, and as
$\hat D_2$ is the harmonic oscillator operator in the $\gamma-$degree of
freedom, we can associate $n$ with the number of phonons in the 
$\gamma-$variable, $n=n_\gamma$.

\section{Solution of coupled equations,
quasidynamic symmetry, quantum numbers and band structure}
\label{s1}\label{rules}
\begin{figure}[!t]
\epsfig{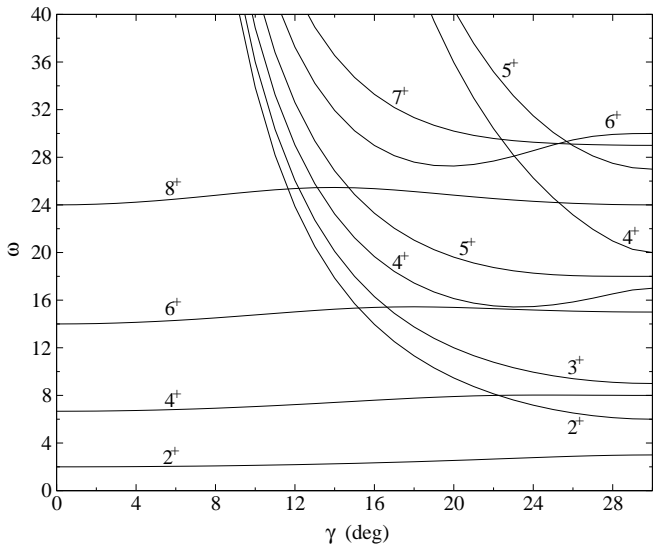}
\caption{Eigenvalues of Eq. (\ref{coupled}) as a function of the asymmetry for
$n_\gamma=0$ (relative to the lowest eigenvalue). In this case the 
solution of the rotational part for the $\gamma-$soft triaxial rotor 
corresponds to the outcome of the Davydov model \cite{Davy}. In addition, 
the soft model
has infinitely many other copies (one for each possible value of $n_\gamma$) 
of these sets of bands at higher energies (depending, of course, on the 
magnitude of $C$).}
\label{rig}
\end{figure}

We discuss here the quantum numbers that appear when solving the 
problem presented in the previous section.
The coupled set of equations (\ref{coupled}) contains the rotational
and $\gamma-$vibrational structure through the functions
$a^L_{K'}$  and the operator $\hat D_2$ respectively. 
In the absence of $\gamma-$vibrations ($n_\gamma=0$), we are left with the 
rotational structure (see Fig. \ref{rig}), which can be analyzed as following.
For each $L$ there is a set of allowed $K$'s: $\{K_i,...,K_f\}$  where 
$K_i=0$ and $K_f=L$ for $L$ even (thus we have $L/2+1$ coupled equations)
and $K_i=2$ and $K_f=L-1$ for $L$ odd (giving $(L-1)/2$ equations).

For $\gamma=0$ and $\gamma=\pi/3$ the 
projection of the third component of the angular momentum on the 
intrinsic axis 3 gives a good quantum number ($K$), while for $\gamma=\pi/6$
the eigenvalue of the projection on the intrinsic axis 1 is a good 
quantum number ($R$). In the intermediate regions, none of them may be taken 
as a good quantum number. 
In the expansion (\ref{expa}) the introduction of $K'$ as a label to
distinguish between the multiple occurrence of states with the same 
value of $L$ is justified because it is referring to the rotational states.

It has been observed that in the triaxial region, moving from $\gamma=30^o$ 
toward $\gamma=0^o$, different groups of states may well
be classified into bands: a first band ($0^+,2^+, 4^+,...$) tends to the 
finite axial rotor values and corresponds to $K=0$; a second band 
($2^+,3^+,4^+,...$) is identified by its behaviour when $\gamma\rightarrow 0$ 
(in Fig. \ref{rig} this group of states somewhat
cluster around $\gamma\sim 10^o-12^o$); the beginning of a third band 
($4^+,5^+,....$) is seen to escape to infinity at a quicker pace (leaving
Fig. \ref{rig} at around $\gamma\sim 20^o$).

The experimental observation that a classification in $\beta$ and $\gamma$
bands seems an almost universal feature of nuclear spectra reinforces
this choice. The labeling with the $K$ quantum number is often encountered 
in the literature, although for what we have said here it may not be
considered adequate. We will in the following use the notation $K^*$ to
regroup the various eigenstates into bands, which are the counterpart of the 
bands with good $K$, that are present in the axial cases.

For phonons with total angular momentum $2$, one usually takes into 
consideration two projections $\mid K\mid =0,2$. 
These projections correspond to the so-called
$\beta-$ and $\gamma-$phonons, that generate axial and non-axial 
quadrupole oscillations, respectively. Because of the non-axial character of 
the $\gamma-$variable, the $\gamma-$vibrations are naturally incorporated 
into Eq. (\ref{coupled}). This is obtained through the operator $\hat D_2$,
which adds the $\gamma-$vibrations to the basic structure of rotations of
a rigid triaxial body. Consequently, one can construct $\gamma-$bands
that consist of copies of the structure of the rigid triaxial rotor (Fig.
\ref{rig}), where every copy possesses a different number of 
$\gamma-$vibrations, characterized by $n_\gamma$. Once the 
rotational and vibrational structure is determined, the eigenvalue $\omega$
can be calculated (in the standard way) and plugged into Eq. \ref{be} to
solve for the $\beta-$part of the problem. An elaborate discussion on the 
$\beta-$part and the chosen potentials will be addressed in section VI.

One may describe the present situation in terms of a quasidynamical symmetry 
\cite{RoL,RoWi} of a peculiar character: at $\gamma=0^o$ the group
SO(2) is a symmetry of the system, associated with $K$, while at $\gamma=30^o$ 
another SO(2) group is a symmetry of the system, associated with $R$, being
the chain U(5)$\supset$SO(3) common to the whole sector $0\le\gamma\le \pi/6$.
In the intermediate region, $0<\gamma< \pi/6$, the SO(2) symmetry is broken,
but it must be noticed (see Fig. \ref{rig}) that the structure of the 
rotational spectrum, present at $\gamma=0^o$, {\it persists} in the whole 
sector without being altered in a dramatic way. Only a smooth and slight 
change may be seen. On the other side the structure of the 'maximally' 
triaxial rotor at $\gamma=30^o$ {\it persists} also in the region around 
$\gamma\sim 20^o-30^o$. In the intermediate region these groups of states
escape to infinity. It must be further noticed that the region where 
the various states, that originate from the axial rotor side, are 
most affected is exactly the region where the states coming from the 
$\gamma=30^o$ triaxial rotor diverge. The strange character of this 
quasidynamical symmetry is that (at variance with the cases 
discussed by Rowe and collaborators \cite{RoL,RoWi,RoI},
where a true phase transition was present between two exactly solvable 
limits associated with different symmetries and different group structures) 
here we are dealing with a smooth transition between two limits which 
formally have the same underlying group structure, SO(2), and there is no 
critical point in between.
Therefore we conclude that the proposed labeling, that retains 
the formal division in $\beta$,$\gamma$ and $K$ 
bands typical of an axial rotor, 
is not only justified by the empirical observation that non-axial nuclei 
display the same classification in bands, but it is also justified in 
view of arguments based on a group theoretical approach. 
It is not clear, at present, if a quantization procedure around a tilted 
axis may help to shed light on this aspect.

\section{Examples}
\label{exam}
Equation (\ref{coupled}) is solved here in a few cases. Applying the recipe 
discussed in Sec. \ref{s1}, the set of differential equations is turned in a 
single algebraic equation in $\omega$. When $L=0$, only the value 
$K'=0$ is present and therefore Eq. (\ref{coupled}) reduces to
\be
\Bigl(\hat D_2-\omega +\sum_i A_i \langle 0\,0\,0\mid
\hat L_i^2\mid 0\,0\,0 \rangle \Bigr) a_0^0(x)=0\,,
\ee
where all the matrix elements are evaluated to be zero and the first 
solution is thus $\omega_{L=0,K^*=0,n_\gamma=0} =\sqrt{C}$
and we may write, in
general, $\omega_{L=0,K^*=0,n_\gamma} =\sqrt{C}(1+2n_\gamma)$.

States with $L=1$ are not present in this model.

For $L=2$, the two possibilities are $K'=0,2$, corresponding to 
 2 coupled equations, 
\begin{eqnarray}
(\hat D_2-\omega+3A_1+3A_2)a^2_0+\sqrt{3}(A_1-A_2)a^2_2=&0 \nonumber\\
\label{ew1} \\
(\hat D_2-\omega+A_1+A_2+4A_3)a^2_2+\sqrt{3}
(A_1-A_2)a^2_0=&0 \nonumber\,,\\
\label{ew2}
\end{eqnarray}
whose solutions are, $ \omega_{L=2,K^*,n_\gamma} =$
\be
\sqrt{C}(1+2n_\gamma)+2\sum_iA_i\pm 
2\sqrt{\sum_i A_i^2-\sum_{i<j} A_iA_j} \,.
\ee
The rotational parts of these expressions reduce to 
the correct values, $3$ and $6$ respectively, when $\gamma=30^o$.

For $L=3$ the only possibility is $K'=2$
and the solution becomes
\be
\omega_{L=3,K^*=2,n_\gamma} =\sqrt{C}(1+2n_\gamma)+4A_1+4A_2+4A_3\,.
\ee
Notice that when $\gamma=30^o$ two of the three components of the 
moment of inertia ($A_2$ and $A_3$)
are equal to $1$ and the remaining ($A_1$) is $1/4$ so the rotational part
of the energy becomes $9$ (as obtained in refs. \cite{LF} and \cite{MTV}).
In general we have
$\omega_{L=3,K^*=2,n_\gamma} =\sqrt{C}(1+2n_\gamma)+ 4\sum A_i$.

For $L=4$ we can write the determinant of the matrix mentioned in Sec. 
\ref{s1} as a third degree equation in $\omega$.
Its three solutions, corresponding to the cases $K^*=0,2,4$, may be 
found analytically, although their expressions are rather lengthy.

It is possible to write an algebraic equation in $\omega$ for every value 
of $L$, but this equation may be solved analytically only for the lowest 
values. We must therefore resort to numerical computation for high values 
of the angular momentum.

In Fig. \ref{rig} we plot the value of $\omega$ for various states for 
$C=0$ with irrotational moment of inertia. This case correspond to the 
well-known rigid rotor solution, that is a particular case of our model
(see Ch. 9 of Ref. \cite{PB}, for example).

Finally we notice that, as a consequence of the definitions given above,
 the following relation holds:
\be
\omega_{200}+\omega_{220}=\omega_{000}+\omega_{320}.
\ee
This expression is a generalization of the well-known relation 
$E(2^+)+E({2'}^+)=E(3^+)$ \cite{PB}.

\begin{figure}[!t]
\epsfig{file=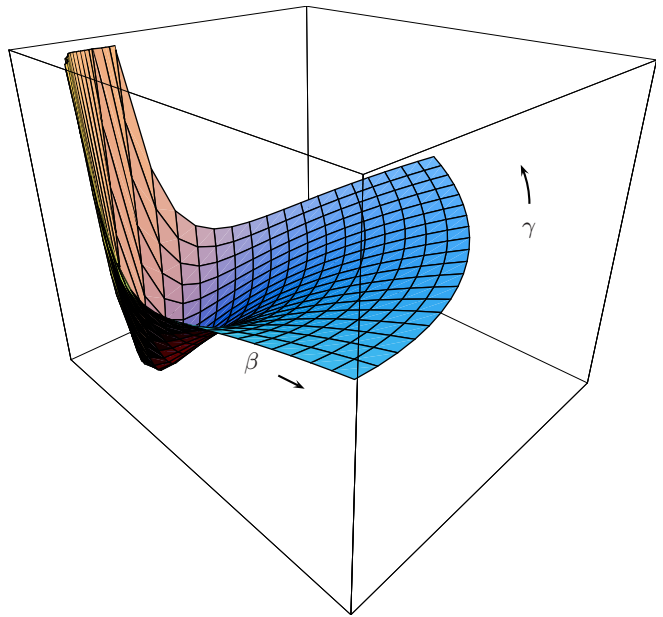,width=0.48\textwidth,clip=}
\caption{(Color online). 
Plot of $u(\beta,\gamma)$ as in formula (\ref{pote}) for 
$0\le \gamma \le \pi/3$. The potential $u_1(\beta)$ has a Kratzer-like form, 
while the potential $u_2(\gamma)$ has a displaced harmonic dependence centered 
around $\pi/6$ (for purpose of illustration).}
\label{polare}
\begin{picture}(1,1)(0,0)
\psset{unit=1.pt}
\rput(-30,160){$\beta$}
\psline{->}(-20,155)(-10,150)
\psarc{->}(40,220){35}{0}{25}
\rput(75,210){$\gamma$}
\end{picture}
\end{figure}

\section{Solution of the $\beta-$part}
Once the $\gamma-$part is solved with a particular choice for the moments of
inertia, we can determine the solution of the $\beta-$part inserting
the appropriate values for $\omega$ in Eq. (\ref{be}). With the substitution 
$\xi(\beta)=\chi(\beta)\beta^{-2}$, Eq. (\ref{be}) is simplified to its
standard form:
\be
{\partial^2 \chi(\beta)\over \partial \beta^2}+\Bigl\{\epsilon-u_1(\beta)
-{(2+\omega)\over \beta^2} \Bigr\} \chi(\beta)=0\,.
\ee
As shown in \cite{LF,FV,FV2} an interesting case is represented by
the Kratzer-like potential (which reduces to a Coulomb-type potential
when $B=0$)
\be
u_1(\beta)=-A/\beta+B/\beta^2\,.
\label{u1}
\ee 
We depict in Fig. \ref{polare}, as an illustration, the surface
corresponding to the reduced potential $u(\beta,\gamma)=u_1(\beta)+
u_2(\gamma)/\beta^2$ where the potential $u_1(\beta)$ has a Kratzer-like form, 
while the potential $u_2(\gamma)$ has a displaced harmonic dependence centered 
(as an example) around $\pi/6$. The depth of the pocket centered around the 
minimum in $(\beta_0,\gamma_0)$ may be adjusted with the parameters of the 
two potentials, $A$, $B$ and $C$.

Following \cite{LF} we can write the solution to equation (\ref{be}) 
directly as:
\be
\epsilon(n_\gamma,n_\beta,L,K^*)={A^2/4 \over \Bigl(\sqrt{9/4+B+
\omega_{L,K^*,n_\gamma}} +1/2+n_\beta \Bigr)^2}\,,
\label{spec1}
\ee
where the values of $\omega_{L,K^*,n_\gamma}$ have been found analytically or
numerically as discussed in the previous section.
Energies are usually redefined fixing the ground state to zero
and using the energy of the first $2^+$ state as unit, namely
\be
\bar \epsilon(n_\gamma,n_\beta,L,K^*) = {\epsilon(n_\gamma,n_\beta,L,K^*)-
\epsilon(0,0,0,0)\over \epsilon(0,0,2,0)-
\epsilon(0,0,0,0)} ~~.
\label{spettro}
\ee
Another interesting case that also leads to a solvable differential equation
is the Davidson potential \cite{Elli,Elli2,RoBa}: 
\be
u_1(\beta)=A_D\beta^2+{B_D \over \beta^2}\,.
\ee
This is an extension of the harmonic potential (that can be obtained when 
$B_D=0$).
Inserting this potential in Eq. (\ref{be}) gives formally the Laguerre 
differential equation and the spectrum may be written as
\be
\epsilon_D(n_\gamma,n_\beta,L,K^*)=\sqrt{A_D} \Bigl( 2n_\beta +
\widetilde \tau_{L,K^*,n_\gamma}+5/2\Bigr) \,,
\label{spec2}
\ee
where, dropping the indices for simplicity, 
$\widetilde \tau$ is the solution of $\widetilde \tau
(\widetilde \tau+3)=B_D+\omega$. The redefinition in 
reduced energy units takes in this case a very compact form:
\be
\bar \epsilon_D(n_\gamma,n_\beta,L,K^*) = {2n_\beta+\widetilde 
\tau_{L,K^*,n_\gamma}-\widetilde \tau_{0,0,0}\over \widetilde 
\tau_{2,0,0}-\widetilde \tau_{0,0,0}  }. 
\ee
The combination of the results for the $\gamma-$part in Sec. 
\ref{s1}-\ref{exam} and the results for the $\beta-$part yields
spectra with an ample range of different possible behaviours.

Another interesting case is inspired by the E(5) and X(5) solutions 
\cite{Iac1,Iac2}.
Starting again from Eq. (\ref{be}) one can adopt a different substitution,
 namely $\phi(\beta)=\beta^{3/2}\xi(\beta)$, and a change of variables,
$z=\sqrt{\epsilon}\beta$. These transformations, together with the choice 
of the 
potential as an infinite square well, yield the Bessel differential equation
\be
\phi''+{\phi'\over z} +\Bigl(1-{\omega+9/4\over z^2} \Bigr)\phi=0\,.
\ee
The solution of the equation above is given in terms of Bessel J functions
of irrational order
\be
\phi_{s,\omega}(\beta)=c_{s,\omega} J_{\sqrt{\omega+9/4}}\Bigl(x_{s,\omega}
{\beta\over\beta_w}\Bigr)\,,
\label{wfnsq}
\ee
where $c_{s,\omega}$ are normalization constants (given analytically in 
\cite{compe}). 
The boundary condition at the wall of the potential well, $\phi(\beta_w)=0$,
implies that the spectrum is given by
\be
\epsilon_{SQ}(n_\beta,n_\gamma,L,K^*)=\Bigl({x_{s,\omega}\over \beta_w}
\Bigr)^2\,.
\ee 
Here and in Eq. (\ref{wfnsq}), $x_{s,\omega}$ is the s-th zero of the 
Bessel function with index that depends on $\omega_{L,K^*,n_\gamma}$. 
Notice that $n_\beta=s$.
The reduced spectrum takes the following form
\be
\bar \epsilon_{SQ}(n_\beta,n_\gamma,L,K^*)={x_{s,\omega_{L,K^*,n_\gamma}}^2-
x_{0,\omega_{000}}^2\over x_{0,\omega_{200}}^2-
x_{0,\omega_{000}}^2}\,.
\ee 
This solution is somewhat similar to the so-called Z(5) solution \cite{BonZ5}
where an infinite square well in $\beta$ was combined with a harmonic 
oscillator centered around $\gamma=30^o$. Major differences are the choice of
exact separation of variables made here, the extension of the solution to 
the whole sector $0^o<\gamma<60^o$ and the possibility to
relax the hypothesis of
irrotational motion in deriving for the components of the moments of inertia.

\section{Random walk fitting procedure}
We are now equipped with a procedure to calculate the eigenvalues of 
the $\gamma-$part of the problem and we have given a few analytic solutions
(Coulomb and Kratzer, harmonic and Davidson, and infinite square well) for
the $\beta-$part. 
The energy spectrum (\ref{spec1}) depends formally on six parameters, 
$A,B,C,A_1,A_2,A_3$, three of which come from the potentials (\ref{u1})
and (\ref{u2}), and the other three are the components of the moment 
of inertia.
A closer look to Eq. (\ref{spettro}) reveals that, once the eigenspectrum
is scaled in the standard way, it does not really 
depend on $A$, while (\ref{spec1}) does. We can now follow the strategy 
to keep the irrotational hypothesis and, since the various 
components of the moments of inertia are connected to each other by means 
of the relations in (\ref{a-moment}), one has to deal with just one component 
(or alternatively with $\gamma_0$). The other two components may be 
thus determined inverting Eq. (\ref{a-moment}):
\be
\gamma_0=arcsin\Bigl({1\over 2\sqrt{A_3}} \Bigr)\,,
\label{gafi}
\ee
and, substituting the value of $\gamma_0$ in the definitions of $A_1$ 
and $A_2$, we obtain the results
\be
A_1\= {4A_3\over \Bigl(\sqrt{12A_3-3}+1\Bigr)^2}\,, \qquad
A_2\= {4A_3\over \Bigl(\sqrt{12A_3-3}-1\Bigr)^2}\,.
\ee
Therefore our model, using the Kratzer potential, has only three parameters, 
$B$, $C$ and $A_3$, that may be used to fit experimental energy spectra.

This reduction of parameters, though convenient, may be very restrictive 
(see \cite{Wood}) and we prefer to implement also a procedure to keep
all the components as independent parameters, at the price of complicating
the fit to experimental data. We distinguish the fitting procedures
with the words irrotational (3 parameters) and non-irrotational fit (5 
parameters).

Similar considerations apply to the energy spectrum of the harmonic/Davidson
potential. In Eq. (\ref{spec2}) six parameters are formally present,
but using the reduced spectrum and the relations among the components
of the moment of inertia, one is lead to a dependence on three of them only
($B_D$, $C$ and $A_3$). Alternatively the non-irrotational fit includes 5 
parameters ($B_D$, $C$, $A_1$, $A_2$ and $A_3$).
Likewise, for the square well case, (4) 2 parameters are
present in the (non-)irrotational fit.

Due to the number of parameters we have preferred to use a numerical method
based on a random walk procedure in order to determine the energy spectrum.

As a first step we consider an isotope and a 
subset of its experimental levels (usually quite small, 4 levels  
typically). 
Starting from some initial set of parameters (typically
$B=C=20$ and $A_3=1$), we minimize the deviation of the 
calculated and experimental energy values by walking
randomly in a suitable part of the many-dimensional parameter space.
The random walk is initially constrained to take only irrotational
moment of inertia (i.e. $A_1$ and $A_2$ are calculated directly from
$A_3$) and consists of a coarse-grain and a fine-grain stage, each of 
which takes a certain number of steps.
The coarse and the fine phases of the procedure differ in the 
percentual amount of the initial parameter values which is allowed 
to change randomly (50\% and 10\% typically).
We usually end up with a set of parameters that correspond to some 
local minimum of our hypersurface (although it must be noticed that 
in this way we are not sure to catch the absolute global minimum).
From this set one may calculate a spectrum, which can be compared
with the experimental one and a total error, $\sqrt{\sum_i (1-E_i^{th}/
E_i^{exp})^2}$, 
that gives an idea of the accuracy of the description that one
may get. Usually, using as input values energies of the ground and
$\gamma-$bands, the typical accuracy may range from 1\% to 0.1\%.
At this point one is still allowed to identify a value of $\gamma_0$, 
calculated from the irrotational moments of inertia (\ref{gafi}), with 
the angular position of the minimum.
As a final step we start from the set of parameters resulting from the 
irrotational fit, but now we relax the irrotational hypothesis, treating
all the components of the moment of inertia as individual parameters.
Often the higher freedom of this fit reduces the error by a factor 
2-10, but sometimes this procedure does not improve substantially 
the quality of the fits. 
It has to be stressed that at this point the identification with a 
single value for $\gamma_0$ does not make sense anymore.

Our goal is, from one side, to give a simple model that goes beyond 
the estimate of ground state energies and furnishes a decent
description of ground and $\gamma-$bands, and, from the other side,
allows one to extract a possible behaviour of the potential energy surface
associated with a given isotope. 

One must keep in mind that this model
will work only when the potential energy surface has just one minimum.
The fact that, after fitting, some important discrepancies result between
calculated and experimental values may point out that a geometrical 
model may not be well applicable after all.

\begin{figure}[!t]
\epsfig{file=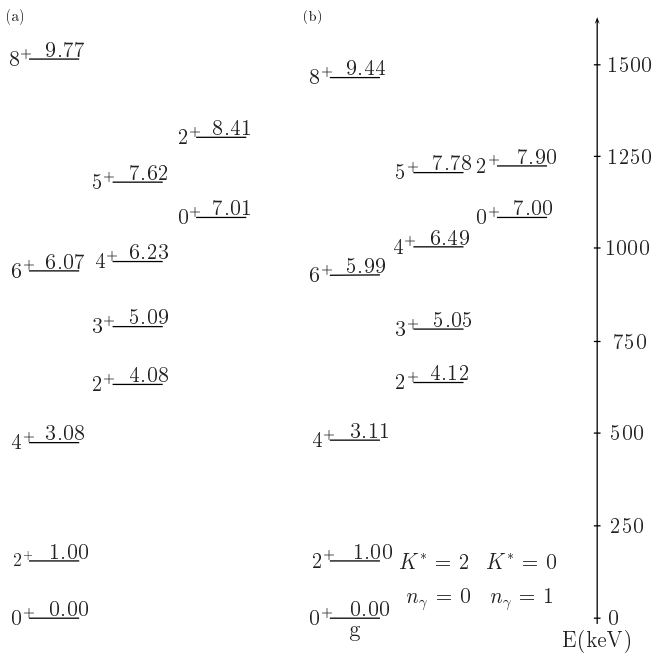,clip=,width=0.48\textwidth}
\caption{Experimental spectrum of $^{188}Os$ (a) and irrotational fit (b)
obtained with $B_D=62.301, C=82.66, A_3=2.3917$. Experimental data
are taken from \cite{NDS}.}
\label{188Os}
\end{figure}

\begin{figure}[!t]
\epsfig{file=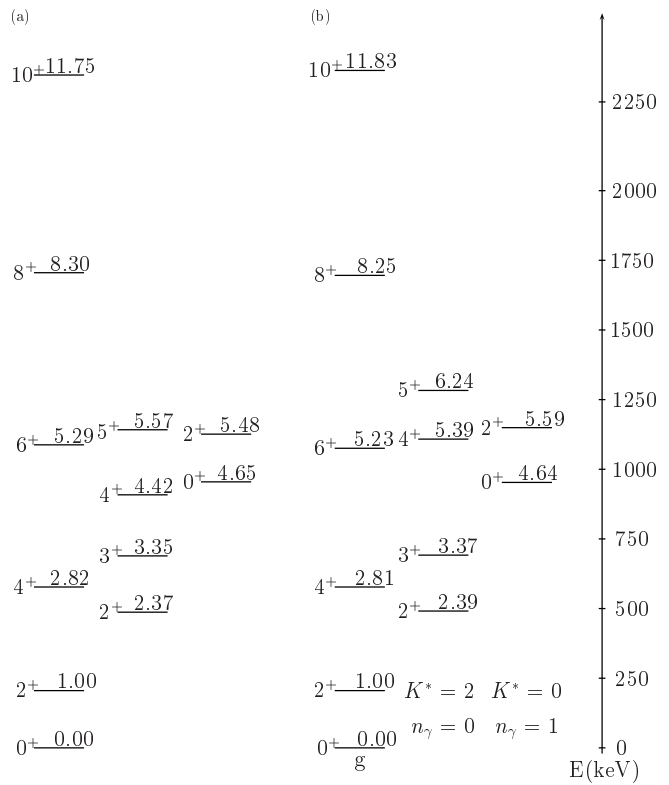,clip=,width=0.48\textwidth}
\caption{Experimental spectrum of $^{192}Os$ (a) and irrotational fit (b)
obtained with $B_D=132.182, C=44.6504, A_3=1.39954$. Experimental data
are taken from \cite{NDS}.}
\label{192Os}
\end{figure}

\section{The osmium isotopes}
\begin{figure}
\epsfig{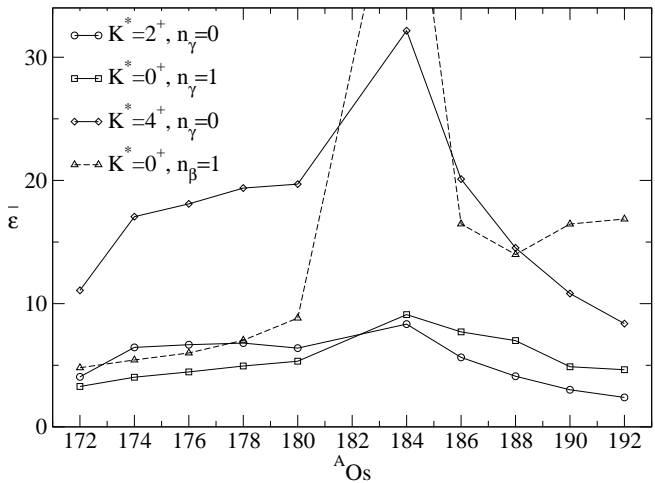}
\caption{Reduced energies of the bandheads of the $(K^*,n_\gamma,n_\beta)=$ 
$(2,0,0)$, $(0,1,0)$, $(4,0,0)$ and $(0,0,1)$ bands (indicated with circles, 
squares, diamonds and triangles respectively)as obtained from
the irrotational fit procedure for the whole chain of osmium isotopes.}
\label{bh}
\end{figure}

We applied the procedure explained in the previous section to the case
of the osmium isotopes with $A=172-180,184-192$.
The irrotational fit with the Davidson potential, starting from the 
$4^+$ and $6^+$ levels of the ground state band and the $2^+$ level of the 
$K^*=2,n_\gamma=0$ band and the $0^+$ level of the $K^*=0,n_\gamma=1$ band, 
leads very quickly to a set 
of parameters ${B_D,C,A_3}$ that gives always a rather good description 
of the low-lying
positive parity energy spectrum, with an estimate on the total error, 
as defined above, which is in the range $0.01-0.09$. In figs. 
\ref{188Os},\ref{192Os} 
we compare the experimental energy levels of $~^{188,192}$Os (a) with the  
irrotational fit (b), obtained from 1600 runs of coarse grained random walk 
plus 1600 runs of fine grained random walk. In this case the non-irrotational
fit does not improve in a sizable way the quality of the fit.
 The absolute error in the case of the 
$10^+$ state is $\simeq 100~$keV, or less. 
In the best cases the accuracy for the $8^+$ and $10^+$ members of
the ground state band may reach the value $1/1000$.
It is observed that the ground-state and excited bands are fairly well
reproduced (which is not surprising since we used the positions of two
levels of each band as an input to our calculations). A bit more surprising
is the qualitative behaviour of other bands: usually we find that the 
$(K^*=0,n_\gamma=1,n_\beta=0)$ and $(K^*=4,n_\gamma=0,n_\beta=0)$ bands 
lie lower (in a few cases dramatically lower) than the first 
$\beta-$band $(K^*=0,n_\gamma=0,n_\beta=1)$. In Fig. (\ref{bh}) we report the
reduced energy of the bandheads of the lowest bands for the whole chain.
In the osmium nuclei with lower mass number, 
we find comparable energies 
for the $\beta-$band and for the $K^*=0$ band, while the $K^*=4$ lies
at considerably higher energies. This is reflected in the magnitudes 
of the $B_D$
and $C$ coefficients: for $^{192}$Os they are respectively, $\sim 130$
and $\sim 44$, but as soon as we move to lower mass numbers the value of $B_D$
drops to less than $1$ for the lightest ones. Thus, the $\beta-$vibrations
are found at much lower energies.

The example displayed in Fig. \ref{188Os} is rather rigid and the 
(dimensionless) values of the parameters
are $B_D=62.3$, $C=82.66$ and $A_3=2.3917$. The latter corresponds
to a value of $\gamma_0\simeq 18.86^o$. The energy difference
between the minimum of the harmonic oscillator and the values at one of the 
borders of the $\Delta\gamma=60^o$ wedge is roughly
\be 
{\hbar^2\over 2 B_m} C {(\gamma-\gamma_0)^2 \over \beta_0^2}\sim 
{0.44\over \beta_0^2} MeV\,,
\label{Ceq}
\ee
which, for a reasonable value of $\beta_0$ (e.g. values 
smaller than 0.3), 
is higher than the excitation energy of the higher lying states shown in Fig. 
\ref{188Os}: $E_{8^+}\sim 1.5 MeV$. 
In order to calculate the value in Eq. (\ref{Ceq}) we used the Bohr-Mottelson 
prescription \cite{BM} for the mass coefficient,
$
B_m={1\over \lambda}{3\over 4\pi}A M_N R_0^2\,,
$
where $\lambda=2$, $M_N$ is the nucleon mass and $R_0=r_0A^{1/3}$ 
is the nuclear radius. Therefore we obtain
\be
{\hbar^2\over 2B_m}={4\pi\hbar^2c^2\over3M_Nc^2r_0^2A^{5/3}}\sim 
{120.72 MeV\over A^{5/3}}\,.
\ee
This expression, combined with the proper mass number, has been used in
Eq. (\ref{Ceq}).
The evaluation of the depth of the potential well gives an indication 
that the wavefunctions of the states that we have fitted are
all well confined inside the well.
In fact the true shape of the potential is thought to depend on 
$\gamma$ periodically (terms like $\cos{3\gamma}$ are the most important),
therefore the harmonic oscillator shape can only represent a convenient
approximation which is valid around the minimum. 

\begin{figure}[!t]
\epsfig{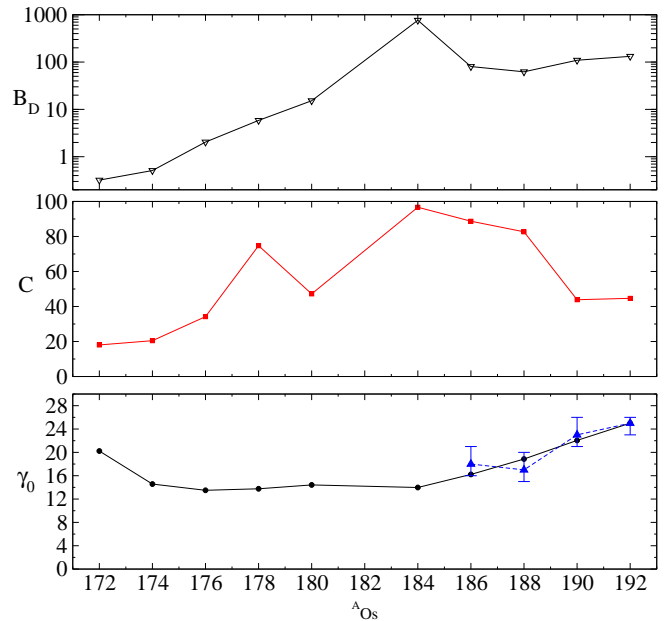}
\caption{Values of $\gamma_0$ from irrotational fit to various osmium
isotopes (line with circles, lower panel) and corresponding 
values of the parameter $C$ (middle panel) and $B_D$ (upper panel). 
Predictions obtained in \cite{Werner} with a different method 
are also shown (dashed line, lower panel).}
\label{gzero}
\end{figure}

We show in Fig. \ref{gzero} the values of $\gamma_0$ (black dots)
obtained from the irrotational fit for the whole chain of osmium isotopes 
that we have analyzed together with the 
corresponding values of the parameter $C$ (red squares, middle part)
and $B_D$ (diamonds, upper part). We notice that,
while the value of $\gamma_0$ decreases from the left to the middle and
then increases, the value of the parameter $C$ does not show any regular 
behaviour with the mass number. When one considers the extremes of the mass 
chain, the rotor becomes softer.
By applying relation (\ref{Ceq}) to the softest case ($^{172}Os$) 
one can check that the harmonic approximation is still valid.

\begin{figure}[!t]
\epsfig{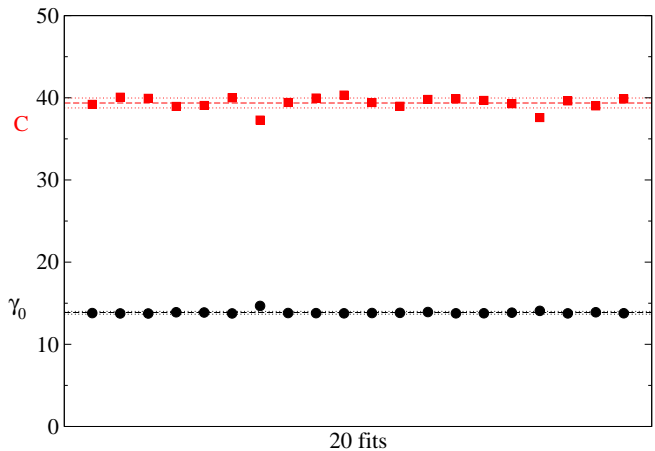}
\caption{(Color online). 
Values of $\gamma_0$ (black dots) and of the parameter C (red 
squares) extracted from 20 independent fits to $~^{178}$Os. 
From these data one obtains an estimate of the theoretical
error associated with the fitting procedure described in the text. 
Mean values, $\langle \gamma_0 \rangle$,
and mean values $\pm$ the corrected standard deviation,
$(\langle \gamma_0 \rangle\pm \sigma_{\langle \gamma_0 \rangle})$, 
are indicated with dashed and dotted lines for both quantities, 
respectively.}
\label{accu}
\end{figure}

In Fig. \ref{gzero}, we also display (dashed line with blue triangles)  
predicted effective values of $\gamma$ using $K$ shape invariants 
\cite{Werner}. Despite the different technique used to obtain those 
values (which are extracted using quadrupole matrix elements and $B(E2)$ 
values) they agree fairly well with our results. For these values, the 
error bars result from errors on the measurement of the transition
rates. In our case an estimate of the theoretical error is quite difficult:
in principle we can repeat the fitting procedure many times and we can extract
a distribution of values of parameters. To understand if our procedure
is reliable we have repeated the fit on $~^{178}$Os 20 times. We have 
collected the various sets of parameters and we present in Fig. \ref{accu}
the extracted values of $\gamma_0$. It is therefore possible to obtain an 
estimate of the distribution of $\gamma_0$ values and to give mean values and
corrected standard deviations as 
a simple estimate for the theoretical error. 

Hartree-Fock-Bogoliubov calculations have been 
carried out (Ref. \cite{Ansari}) in the Os-Pt region with results pointing
toward a prolate structure with $\gamma_0$ very close to $0^o$ and with a 
flat potential in $\gamma$ (for the Os nuclei  with $A=186-192$). 
In such self-consistent calculations the potential energy surface is 
calculated variationally, starting from a given effective interaction.
In most cases no dynamical calculations are done on top of that and  
energies are often associated with the exact 
value of the minimum of this surface. Likewise, a few excited states can 
also be predicted.
Our approach starts from a given mathematical expression for the potential 
$V(\beta,\gamma)$, which is then used to solve the collective model and can
predict full energy spectra.
However, the particular choice of $V(\beta,\gamma)$ made at present may
still deviate from actual, more realistic, potential energy surfaces.
Besides, indications exist for ground state hexadecapole deformation 
in the Pt-Os region, as suggested by 
strong E4 coupling arising in $(\alpha,\alpha')$ and $(p,p')$ reactions.

Very recently a description of Os isotopes (among others) has been proposed
within the IBM-1 \cite{Mac}. The authors fit the low-energy positive-parity 
spectra obtaining a good overall description. Both their work and ours,
which rely on models based on different ingredients, describe
equally well a good fraction of the spectral properties of Os isotopes,
although a number of differences clearly leave room for more detailed studies.
Another analysis of quadrupole moments, transition and transfer rates 
of (platinum and) osmium isotopes in terms of IBM-2 \cite{bij} shows 
that a reasonably good agreement with experimental data can be obtained with a 
smoothly varying set of parameters. 

In our present study to construct solutions to the Bohr hamiltonian for soft
triaxial nuclei, the model parameters used determine (i) the moments of 
inertia (through the parameter $A_3$, or $A_i$ for the more general fit), 
(ii) the stiffness parameter $C$ for the $\gamma-$oscillatory motion and
(iii) $B_D$ and $A_D$ determining the Davidson potential. There appear rather 
large variations in particular in $B_D$ and $C$; however, the deduced value
of the quantity $\gamma_0$ (see fig. \ref{gzero} - lower part) still 
exhibits a very smooth variation and agrees very well with results using a 
totally different approach \cite{Werner}. We stress that the above variations 
do not at all contradict the fact that smooth variations in the parameters 
for the IBM-2 hamiltonian result in describing the observed data in the Os 
nuclei, because totally different models are used.

\begin{figure}[!t]
\epsfig{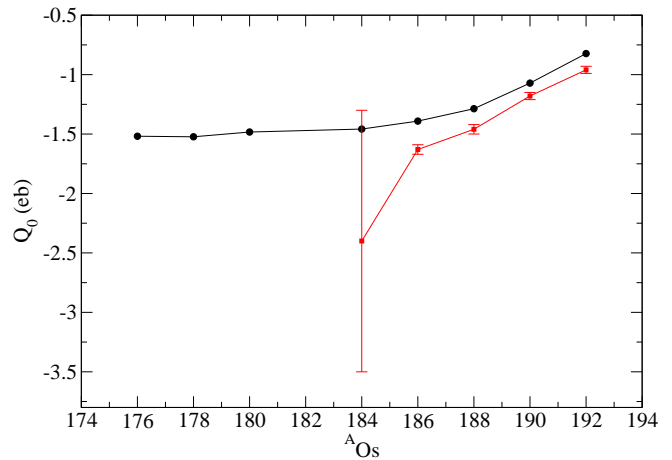}
\caption{(Color online). Electric quadrupole moments for osmium isotopes in 
$eb$. Data (red squares with error bars) are taken from Ref. \cite{Ragh}.
The calculated values (black dots) have been obtained by adjusting an 
additional parameter to reproduce the excitation energy of the first $2^+$ 
state.}
\label{qmom}
\end{figure}

In order to test even further our model, we present in fig. \ref{qmom} a 
comparison of calculated and experimental static quadrupole moments of the
 first excited $2^+$ state. The definition of the quadrupole moment adapted 
to the present case is:
\be
Q=\sqrt{16\pi\over 5} \biggl( {2 \atop 2} {2 \atop 0} {2\atop -2}\biggr)
{3ZR_0^2 \over 4\pi} \langle 2_1^+ \mid\mid \alpha_2^*\mid\mid2_1^+\rangle
\ee
where the reduced matrix element, once we exploit the trigonometrical 
simplification for the $\gamma$ part and we insert the values of the 
Clebsch-Gordan coefficients, is calculated to be
$$
\langle 2_1^+ \mid\mid \alpha_2^*\mid\mid2_1^+\rangle=\langle \xi_0(\beta)
\mid\beta\mid \xi_0(\beta)\rangle {10\over \sqrt{70}} \bigl( 1-
{1\over 4C}\bigr) $$
\be
\bigl[ \cos{\gamma_0} (a_2^2-a_0^2)+2a_0^2a_0^2\sin{\gamma_0}\bigr] \,.
\ee
The last expression still contains the matrix element of $\beta$ that 
depends on the Davidson ground state wavefunction, which in turn contains 
the parameter $A_D$. This parameter can be fixed by requiring that 
$\epsilon_D(2^+)-\epsilon_D(0^+)$ matches with the absolute energy difference 
in MeV. The results collected in fig. \ref{qmom}, are correct both in sign 
and trend, but they underestimate a bit the measured values \cite{Ragh}. 
A more complete study, including $B(E2)$ values, is in progress.

\section{Conclusions}
In the present paper, we have presented a method in order to solve 
the Bohr-Mottelson collective model for a general soft triaxial nucleus.  
By choosing a potential of the form $V(\beta,\gamma)= V_1(\beta) +
V_2(\gamma)/\beta^2$,  the Bohr Hamiltonian separates exactly in the 
$\beta$ and $\gamma$ variables. Using a displaced harmonic oscillator 
potential in the $\gamma$ direction, representing the softness in 
the $\gamma$ variable, allows us to solve approximately the $\gamma$-angular 
equation. In solving that part of the problem, we discuss two 
possibilities to further separate the $\gamma$ variable from the 
projections of the angular momentum. One approximation neglects 
fluctuations of the moments of inertia in the $\gamma$ direction, but 
keeps the softness. Under this assumptions, the three components of the 
moment of inertia are related and result in one parameter. 
The other approximation allows a fitting of the three components of the 
moment of inertia independently. This is in line with the observation 
that the experimental data do not agree so well with the assumption of 
irrotational flow. We then present an algebraic method in order to find 
approximate solutions to the set of coupled equations that result for 
the motion in the $\gamma$ variable and also discuss at some length the 
state labeling problem in order to characterize the various collective 
bands that result. We subsequently study the equation for the $\beta$ 
degree of freedom, which can be solved exactly for a number of interesting 
potentials. We thus consider the Coulomb/Kratzer potential, the Davidson 
potential (which is an extension of the harmonic oscillator potential 
in $\beta$) as well as the infinite square-well potential in our calculations.
In each case, we are able to derive analytic expressions, which are then 
used to determine the full energy spectra. These energy spectra, that 
describe oscillatory behavior in both the $\beta$ and $\gamma$ variables, 
as well as the rotational motion, in general depends on 6 parameters: 
three that characterize the $V(\beta,\gamma)$ potential and three that 
determine the moment of inertia. Using appropriate scaling in the energy 
spectrum only 5 remain to be determined. Going back to irrotational motion, 
the three components of the moment of inertia are related and this reduces 
the full set to just 3 parameters. Due to the number of parameters and the 
very involved energy eigenvalue expression, we have used a numerical method 
based on a random walk procedure in order to obtain the optimal fitting 
parameter set. We have used this method to study the Osmium isotopes in 
the interval $172 \le A \le 192$. One of the results is the fact that the 
equilibrium $\gamma_0$ value shows a rather smooth dependence on $A$, 
but the stiffness of the potential in the $\gamma$ variable indicates
rigid cases for intermediate masses and softer cases for isotopes
sitting at the extremes of the considered mass chain.
To elucidate this point it would be very helpful to search for solutions
of Eq. (\ref{g-a}) with more involved periodic potentials \cite{Stij}.
Still, the calculated values of 
$\gamma_0$ are in rather good agreement with an independent approach 
to extract an effective $\gamma$ value by Werner {\it et al.}.
We have in mind to study the rare-earth region in a more systematic 
way using the methods presented here. 

Recently other works have appeared that treat similar problems, yet
with different spirit and aims. Caprio \cite{CapNew} analyzes the numerical
diagonalization of a $\beta-$soft, $\gamma-$stabilized problem, evidencing
how the approximate separation of variables of the X(5) model may be questioned
on the basis of a strong $\beta-\gamma$ coupling. Despite the fact that
our approach is complementary (rather than solving exactly the numerical
problem, we postulate analytically solvable potentials and we try to see
if they can be profitably used), we reach similar conclusions. For example
we also find that a considerable degree of dynamical $\gamma-$softness is
needed (for realistic cases of $\gamma$ stiffness) to account for the energies
of the $\gamma-$bands' levels. Furthermore our method allows
 to look for the optimal value of the actual moments of inertia. A point of
difference is, instead, the pattern for the staggering: in the examples below
we rather find a scheme of the type $(2^+,3^+),(4^+,5^+),...$ similar to that
of the rigid triaxial rotor. Another very important series of paper by Rowe
and collaborators \cite{Rowe-new,RoI,RoTuRe} furnishes a rapidly
converging method for exact numerical treatment of the problem by using a
basis defined in terms of ``deformed'' wavefunctions in $\beta$ and
five-dimensional spherical harmonics.

~\\
The authors are most grateful to J.Wood for extensive discussions and
L.F. wishes to thank A.Vitturi for profound comments.
Financial support from the "FWO Vlaanderen" (L.F. and K.H.) and the 
University of Ghent (S.D.B. and K.H.), that made this research possible, 
is acknowledged.

\section*{APPENDIX: Matrix elements}
List of non-null matrix elements of operators as discussed in Eq. 
(\ref{undici}) of Sect. II.
\be
\langle LMK\mid \hat L_3^2 \mid LMK'\rangle =K^2 ~~~\mbox{~~if~~} K=K'
\ee

\begin{widetext}
\be
\langle LMK\mid \hat L_1^2 \mid LMK'\rangle =
\left\{
\begin{array}{lr}
\displaystyle
{L(L+1)-K^2\over 2} &~~~\mbox{~~if~~} K'=K\\
\displaystyle
{\langle LMK\mid \hat L_+^2 \mid LMK'\rangle \over 4}&~~~\mbox{~~if~~} K'=K-2\\
\displaystyle
{\langle LMK\mid \hat L_-^2 \mid LMK'\rangle \over 4}&~~~\mbox{~~if~~} K'=K+2\\
\end{array}
\right.
\ee

\be
\langle LMK\mid \hat L_2^2 \mid LMK'\rangle =
\left\{
\begin{array}{lr}
\displaystyle
{L(L+1)-K^2\over 2} &~~~\mbox{~~if~~} K'=K\\
\displaystyle
-{\langle LMK\mid \hat L_+^2 \mid LMK'\rangle \over 4}&~~~\mbox{~~if~~} 
K'=K-2\\
\displaystyle
-{\langle LMK\mid \hat L_-^2 \mid LMK'\rangle \over 4}&~~~\mbox{~~if~~} 
K'=K+2\\
\end{array}
\right.
\ee
where
\be
\langle LMK\mid \hat L_+^2 \mid LM(K-2)\rangle=
\sqrt{(L+K)(L+K-1)(L-K+2)(L-K+1)} 
\ee
and
\be
\langle LMK\mid \hat L_-^2 \mid LM(K+2)\rangle=
\sqrt{(L+K+2)(L+K+1)(L-K-1)(L-K)} ~\,.
\ee
\end{widetext}

\end{document}